\documentstyle[prl,aps,epsf]{revtex}
\begin{document}

\twocolumn[
\hsize\textwidth\columnwidth\hsize\csname@twocolumnfalse\endcsname

\title{Variable electrostatic transformer: controllable coupling of
two charge qubits}

\author{D.~V. Averin$^1$ and C. Bruder$^2$}
\address{$^1$Department of Physics and Astronomy, University of 
Stony Brook, SUNY, Stony Brook, NY 11794-3800} 
\address{$^2$Departement Physik und Astronomie, Universit\"at Basel 
Klingelbergstrasse 82, 4056 Basel, Switzerland} 

\date{\today}

\maketitle

\begin{abstract}
We propose and investigate a novel method for the controlled coupling 
of two Josephson charge qubits by means of a variable electrostatic 
transformer. The value of the coupling capacitance is given by the 
discretized curvature of the lowest energy band of a Josephson junction, 
which can be positive, negative, or zero. We calculate the charging 
diagram of the two-qubit system that reflects the transition from 
positive to negative through vanishing coupling. We also discuss how 
to implement a phase gate making use of the controllable coupling.
\end{abstract}

\pacs{PACS numbers:74.50.+r}

]

Following experimental demonstrations of individual qubits implemented
with Josephson junctions operated in the charge \cite{b1,b2} or flux
\cite{b3,b4,b5,b5*} regimes, a lot of interest is now focused
on building multi-qubit Josephson circuits.  Recently, there has been
an encouraging first experimental step \cite{b6} demonstrating
quantum-coherent dynamics of two charge qubits coupled directly
through a capacitance. The coupling strength could not be varied in
this experiment, since there is no simple way of changing the
electrostatic capacitance between two metallic islands. A controlled
coupling of charge qubits would be highly desirable. Many
two-qubit transformations that could be implemented with a
controllable coupling become more difficult or impossible at constant
coupling, since in the presence of interactions one needs to correct
for different dynamic phases of the two-qubit states. Two-qubit gates
with untunable couplings have been suggested \cite{b6*}, however, the
gate design becomes significantly more complex.

For flux qubits, the problem of a controlled coupling can be
solved with variable flux transformers (see, e.g., \cite{b7}) which
can be implemented naturally with the standard tools of superconductor
electronics \cite{b8}. For other types of ``magnetic'' qubits, e.g.,
spin qubits, the possibility to control the qubit coupling both in
sign and absolute value has been predicted theoretically
\cite{burkard} (for recent experimental progress, see \cite{bg}). For
charge qubits, however, a comparably natural solution has been lacking
although the importance of the interaction control problem has been
recognized for some time. Suggested controlled coupling
methods include direct spatial separation of qubit states by adiabatic
transfer along junction arrays \cite{b9} or coupling through
superconducting resonators \cite{b10,b11,b21,b12,b22,b13,b20}. The 
first one is too complicated to be practical at present. The
second requires relatively large inductances of small geometric 
dimension that are fundamentally difficult to produce, in particular 
without creating additional dissipation at the relevant large 
frequencies.

\begin{figure}[htb]
\setlength{\unitlength}{1.0in}
\begin{picture}(3.,1.2) 
\put(.7,.1){\epsfxsize=1.6in\epsfbox{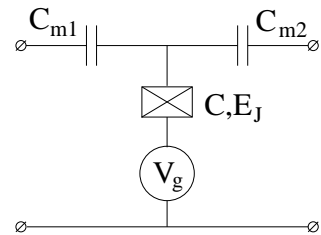}}
\end{picture}
\caption{Equivalent circuit of the variable electrostatic 
transformer for the controlled coupling of charge qubits. 
The Josephson coupling energy $E_J$ of the small tunnel junction 
and the capacitances $C$, $C_{m1}$, $C_{m2}$ of the structure determine 
the dispersion relation of the junction. Its lowest energy band 
$\epsilon_0(q)$ provides the required variable coupling 
capacitance controlled by the gate voltage $V_g$.}
\label{fig1} \end{figure}

The purpose of this work is to analyze a new circuit to
implement a controllable coupling of charge qubits. The circuit, shown
in Fig.~\ref{fig1}, generalizes the simple capacitive coupling used in
\cite{b6} and makes use of the following basic feature of an
individual small Josephson tunnel junction. If the junction dynamics
is confined to the lowest energy band of its band structure, it
behaves as a variable capacitance with respect to the injected charge
\cite{b14}. The magnitude of this capacitance depends on the bias
point and can even be made {\em negative}. This is important for
the application as a controllable coupling element, since it enables
one to drive the coupling capacitance to zero.
The physics of the capacitance modulation is
the transformation of charge, when an injected charge causes a
Cooper-pair transfer across the junction that changes the output
charge. If, however, the two qubits are coupled directly through a
small junction (the coupling scheme considered, e.g., in a somewhat
different context in \cite{siewert,b15}) this mechanism would not
produce a variable-capacitance coupling scheme, since variations of
the junction capacitance are compensated by a redistribution of the
transferred charge on the capacitances in series with the junction.
Nevertheless, if the junction is
included ``perpendicular'' to the coupling direction, as in
Fig.~\ref{fig1}, the required ``variable electrostatic
transformer'' with a gate-voltage-controlled coupling capacitance is produced.

Indeed, if the voltage $V$ is applied to the pair of input terminals 
(left or right, $i=1,2$) of the circuit shown in Fig.~\ref{fig1}, the
Hamiltonian of the system is that of an individual Josephson junction
(or ``Cooper-pair box'' \cite{b16,b17}) with the charge $2e(q-q')$
injected into it, where $q=C_{mi}V/2e$ and $q'=C_{mi}V_g/2e$. 
When the junction is in its lowest energy band $\epsilon_0$, the
output voltage $V_{out}$ at the opposite terminals varies such that
\begin{equation}
4e^2/C_0 \equiv 2e \partial V_{out} / \partial q = \partial^2 
\epsilon_0 (q-q')/ \partial q^2 \, .
\label{e1} \end{equation} 
The input-output interaction strength defined by this relation
varies 
with the charge $2eq'$ injected by the gate voltage and goes 
through zero at a bias point dependent on the shape of the energy 
band $\epsilon_0(q)$. 

\begin{figure}[htb]
\setlength{\unitlength}{1.0in}
\begin{picture}(3.,1.5) 
\put(-.15,.1){\epsfxsize=3.5in\epsfbox{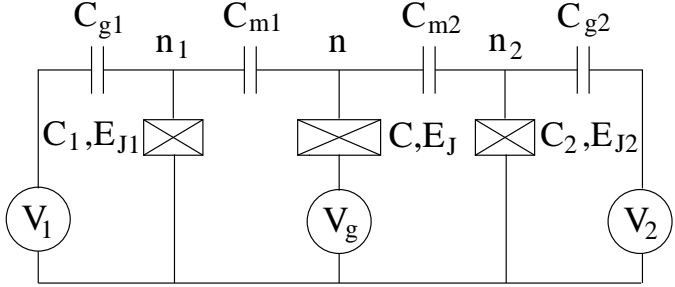}}
\end{picture}
\caption{Equivalent circuit of the two charge qubits coupled by 
a variable electrostatic transformer.}
\label{fig2} \end{figure}

If the transformer 
shown in Fig.~\ref{fig1} 
is inserted between 
two charge qubits (Fig.~\ref{fig2}), it provides a 
gate-voltage-controlled qubit coupling. The qubits 1,2 with 
Josephson coupling energies 
$E_{Ji}$, $i=1,2$, are biased by the gate voltages $V_i$ 
through the capacitances $C_{gi}$ and are coupled to the central 
island of the transformer by capacitances $C_{mi}$. Inverting the 
capacitance matrix of this system, 
we can write the Hamiltonian as
\begin{eqnarray}
H= &-&\sum_{i=1,2}E_{Ji}\cos \varphi_i + E_{Ci}(n_i-q_i)^2
\nonumber\\
&-&E_J\cos \varphi + E_C (n-q(n_1,n_2))^2 \, ,
\label{e2} \end{eqnarray}
where $E_C=2e^2/(C_{\Sigma} -\sum_i C_{mi}^2/C_{\Sigma i})$,
$E_{Ci}=2e^2/C_{\Sigma i}$, 
$n_i$ is the number of excess Cooper pairs on qubit
$i$, and $q_i\equiv V_iC_{gi}/2e$. The qubits
are coupled through the charge $q$ induced on the
transformer junction: $q=q_g-\sum_i (n_i-q_i) C_{mi}/C_{\Sigma i}$,
where $q_g \equiv (V_g/2e)\sum_i C_{mi}(1- C_{mi}/C_{\Sigma
i})$. The Josephson phase differences across the respective
junctions are denoted by $\varphi_i$ and $\varphi$. Finally, 
$C_{\Sigma i} \equiv C_i+C_{gi}+C_{mi}$, and $C_{\Sigma} \equiv
C+C_{m1}+C_{m2}$.

We assume that the degrees of freedom of the transformer junction 
are fast, and that the junction is confined to the lowest energy 
band of its band structure. Then we can replace the part of the 
Hamiltonian related to the transformer by the dispersion relation 
$\epsilon_0(q)$ of this lowest band, 
\begin{equation}
-E_J\cos \varphi + E_C (n-q(n_1,n_2))^2 \rightarrow 
\epsilon_0(q(n_1,n_2))\, .
\label{e3} \end{equation}
This is the analogue of the Born-Oppenheimer approximation which in
our case is valid if the characteristic energy gap between the bands
of the transformer junction is much larger than the qubit
energies. For our qubit coupling, this condition requires that $E_J\gg
E_{Ji}$ for $E_J\ll E_C$, and $(E_CE_J)^{1/2}\gg E_{Ji}$ for 
$E_J\geq E_C$ (the qubits are assumed to be in the charging regime).

To determine the qubit coupling provided by the 
transformer we will use the known properties of the junction 
bandstructure \cite{b14,b18}. Also, we use the fact that in the 
charge qubit regime ($E_{Ji}\ll E_{Ci}$, $q_i\simeq 1/2$), the 
dynamics of the charges $n_i$ in (\ref{e2}) is reduced to two 
states, so that the charges can be expressed through the Pauli 
matrices, $n_i=(\sigma_{zi}+1)/2$. Assuming that the structure in 
Fig.~\ref{fig2} is symmetric, i.e., $C_{m1}/C_{\Sigma 1}= 
C_{m2}/C_{\Sigma 2} \equiv c$,
the transformer term $\epsilon_0(q(n_1,n_2))$ in the 
reduced Hamiltonian can be expressed as follows: 
\begin{equation}
\epsilon_0(q(n_1,n_2))= \nu \sigma_{z1}\sigma_{z2}
 + \delta (\sigma_{z1}+\sigma_{z2}) +\mu \, . 
\label{e4} \end{equation} 
The coupling coefficient $\nu$ is given by 
\begin{equation}
\nu =\frac{1}{4}[\epsilon_0(q_0+c) +\epsilon_0(q_0-c)- 
2\epsilon_0(q_0)] \, ,
\label{e6} \end{equation} 
where $q_0 = q_g+c\sum_i (q_i-1/2)$.
The term linear in $\sigma_z$ shifts the qubit bias by 
$ \delta = [\epsilon_0(q_0+c) -\epsilon_0(q_0-c)]/4$, 
and the constant term $\mu= [\epsilon_0(q_0-c) +\epsilon_0 (q_0+c)+ 
2\epsilon_0(q_0)]$ affects the energy of all two-qubit states and 
is not relevant as long as we are discussing one pair of qubits. 

\begin{figure}[tb]
\setlength{\unitlength}{1.0in}
\begin{picture}(3.,2.2) 
\put(.1,.0){\epsfxsize=2.8in\epsfbox{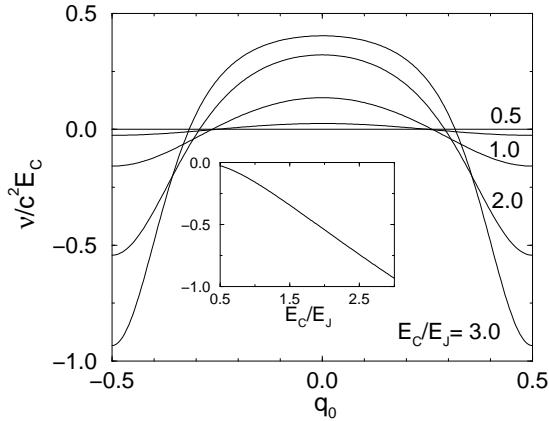}}
\end{picture}
\caption{Coupling energy $\nu$ of the two qubits in units of 
the charging energy $E_C$ of the transformer junction as a function 
of the average induced charge $q_0$. The coupling strength is $c=0.1$, 
and the ratio $E_C/E_J$ is given for each 
curve. The normalization of $\nu$ by $c^2$ makes it independent 
of $c$ for small $c$. Inset: $\nu(q_0)$ at $q_0=\pm 0.5$, as a 
function of $E_C/E_J$ for the same coupling strength $c$.} 
\label{fig3} \end{figure}

Equation \ref{e6} is one of the central results of our work. We see 
that $\nu$ has the
structure of a discretized second derivative of $\epsilon_0$ similarly
to Eq.~(\ref{e1}). Since $\epsilon_0(q)$ is a periodic function,
$\nu$ can be positive or negative. For instance, in the tight-binding
limit $E_C\ll E_J$, when $\epsilon_0(q)=-\Delta \cos (2\pi q)$, the
coupling is:
\begin{equation}
\nu =\frac{\Delta}{2}\cos (2\pi q_0)(1- \cos (2 \pi c))\, , 
\label{e7} \end{equation} 
and $\delta = (\Delta/2)\cos (2\pi q_0)\cos (2\pi c)$. Obviously, 
the coupling can be controlled by the gate voltage through the 
average induced charge $q_0$, and can change sign. In general, 
the coupling constant can be calculated numerically from Eq.~(\ref{e6}). 
The results are plotted in Fig.~\ref{fig3} which shows that in 
accordance with Eq.~(\ref{e7}), the $q_0$-dependence of $\nu$ is 
harmonic for $E_C\leq E_J$, and changes sign at $q_0 \approx \pm 
1/4$. At larger $E_C$, $\nu(q_0)$ becomes non-harmonic, and for 
$E_C\gg E_J$ and small coupling coefficients $c$, approaches $c^2E_C/2$ 
for all $q_0$ except for the vicinity of the point $q_0= \pm 1/2$, 
where $\nu/c^2E_C$ becomes large in absolute value and negative: 
$\nu/c^2E_C \simeq -E_C/2E_J$ for $c\ll E_J/E_C$, and $\nu/c^2E_C 
\simeq -1/2c$ for $E_J/E_C\ll c \ll 1$. The periodic nature of 
$\epsilon_0(q)$ which has a minimum at $q=0$ and a maximum at $q= 
\pm 1/2$ implies that the maximum of the absolute value of $\nu$ is 
reached for $c=1/2$ and is equal to half the bandwidth, e.g., 
$E_C/8$ for $E_C\gg E_J$, and $\Delta$ for $E_C\ll E_J$ -- see 
Eq.~(\ref{e7}).

\begin{figure}[htb]
\setlength{\unitlength}{1.0in}
\begin{picture}(3.,2.6) 
\put(-0.1,.0){\epsfxsize=3.2in\epsfbox{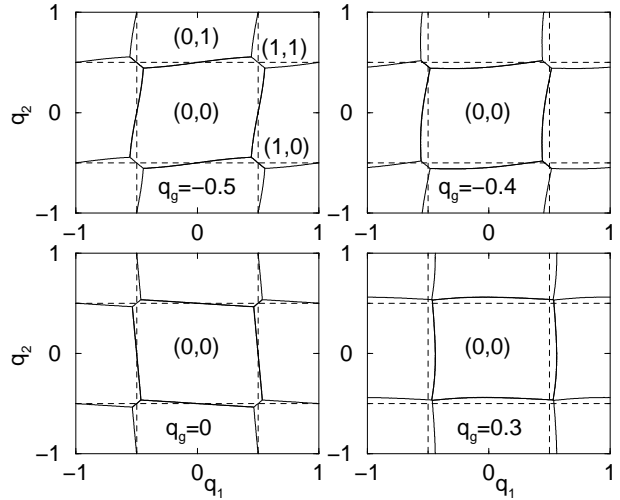}}
\end{picture}
\caption{Charging diagram of two charge qubits coupled 
through the variable electrostatic transformer for $c=0.3$ and 
$E_C/E_J=3$. The curves show the boundaries of the stability regions of 
the charge states $(n_1,n_2)$ in the plane of the two induced charges 
$q_i\equiv V_iC_{gi}/2e$, $i=1,2$. The gate charge $q_g$ is controlled 
by the transformer gate voltage $V_g$. The changing shape of the 
stability region reflects the $V_g$-controlled transition from 
positive to negative effective coupling capacitance of the two qubits.
Dashed lines: uncoupled case.}
\label{fig4} \end{figure}

The controlled electrostatic coupling of two qubits offers new 
possibilities to manipulate two-qubit states. The simplest one is 
the change of the structure of the {\em charging diagram} of the 
transformer-coupled qubits with the gate voltage $V_g$. The charging 
diagram shows the regions of stability of the charge states $(n_1,n_2)$ 
as a function of the induced charges $q_i$, for small Josephson 
coupling $E_{Ji}\ll E_{Ci}$. The diagram is periodic with period $1$ 
in both charges $q_i$, and has a honeycomb structure in the case of 
fixed electrostatic coupling. Figure \ref{fig4} shows examples
of the charging diagram.
The diagram was obtained by minimizing the electrostatic 
qubit energy together with the coupling energy (\ref{e3}). 

At $c=0.3$, the main qualitative features of Fig.~\ref{fig4} can be
understood neglecting the difference between $q_0$ and $q_g$, from the
$q_0$-dependence of the qubit interaction, see Fig.~\ref{fig3}.  For
$q_g=0$, the coupling is positive and the charging diagram has the
usual shape characteristic of a fixed electrostatic coupling. The
cells of the honeycomb structure are tilted in the $q_1=-q_2$
direction, since increasing the gate voltage of one qubit makes it
easier to add a charge to the other qubit. In the absence of coupling,
the degeneracy point of the four charge states $(0,0)$, $(0,1)$,
$(1,0)$, $(1,1)$ is extended in the $q_1=q_2$ direction into the
boundary between $(0,1)$, $(1,0)$, since positive
coupling makes the charging energy of these two states lower than that
of $(0,0)$, $(1,1)$. If the coupling is mostly negative, e.g., for
$q_g= -0.5$, the cells of the honeycomb structure are tilted in the
$q_1=q_2$ direction. The degeneracy point of the four charge states is
changed differently, since the states $(0,0)$, $(1,1)$ now have a
lower charging energy than $(0,1)$, $(1,0)$. The degeneracy point is
extended in the $q_1=-q_2$ direction and creates a direct boundary
between $(0,0)$ and $(1,1)$. Around $q_g= 0.3$, the qubit
coupling is suppressed. Hence, four neighboring stability regions
touch in one point, and their shape is approximately rectangular like
for non-interacting qubits. In contrast to the case of fixed coupling,
all the cells in Fig.~\ref{fig4} are curved, since changing $q_1$,
$q_2$ also changes the effective charge $q_0$ of the transformer
island, so that the effective qubit coupling constant $\nu$ and the
renormalization $\delta$ of the qubit bias change with $q_i$. This
effect is largest for values of $q_g$ around which $\nu(q_0)$ varies
strongly, like e.g., $q_g=-0.4$, see Figs.~\ref{fig3} and \ref{fig4}.

The possibility to switch the coupling between two qubits on and off 
is important for the realization of practically all types of two-qubit
gates. An advantage of our scheme is that the effective qubit coupling 
can be suppressed even in presence of a small direct positive geometric 
capacitance between the qubit islands. This ``parasitic'' capacitance 
only shifts the values of the gate voltages at which the total coupling 
vanishes. As an example, we show how our proposal can be used to directly 
realize a {\em phase gate}, a transformation that changes the sign of the
$|11\rangle$ state and does not change the amplitudes of the other three
two-qubit states. If the qubit tunneling amplitudes are tuned to
zero, the Hamiltonian of the two-qubit system (Fig.~\ref{fig2}) with
coupling (\ref{e4}) is
\begin{equation}
H= \nu \sigma_{z1} \sigma_{z2} + \eta (\sigma_{z1}+\sigma_{z2}) \, ,
\label{e8} \end{equation} 
where $\eta =\delta-(q_i-1/2)E_{Ci}$. (For simplicity, we again 
discuss only the symmetric structure, which for the purpose of 
realizing the phase gate also implies identical qubit charges 
$q_i$. All results of our work can be directly extended to the
asymmetric case.) By adjusting the three gate voltages $V_1$, 
$V_2$, and $V_g$ such that 
$\nu(q_0=q^*)=0$ and $q_i=1/2+\delta(q^*)/E_{Ci}$
one can make the Hamiltonian (\ref{e8}) vanish. 
For $E_J\ll E_C$ such a stationary qubit state is realized when 
$q_i=1/2$ and $q_0=q_g=1/4$. 
By applying gate-voltage pulses one 
can temporarily switch on both the interaction and the qubit 
bias in such a way that the qubit states accumulate dynamic 
phases \cite{b19}. To obtain a phase gate, i.e., to have the state 
$|11\rangle$ accumulate the phase $\pi$ up to an irrelevant 
common phase of all four qubit states, the phase $\phi_{\nu}= 
\int dt \nu(t)$ due to the interaction energy and the phase 
$\phi_{\eta}=\int dt \eta(t)$ due to the qubit 
bias should be chosen as $\phi_{\nu}=\phi_{\eta}=\pi/4$. 

These conditions fix only the time integral of the gate-voltage 
pulses. Their shape can be chosen to reduce the excitation 
amplitude of the upper energy band of the 
transformer junction that would invalidate our Born-Oppenheimer 
approximation (\ref{e3}) and effectively entangle the qubit 
states with the states of the transformer creating an additional 
decoherence mechanism. Transitions to the upper energy bands 
are suppressed if the pulses are sufficiently slow on the scale 
of the transformer energies. In this case, the amplitude 
$\alpha_k(t)$ of an excitation process to the $k$th band can be 
calculated by standard adiabatic perturbation theory as
\begin{equation}
\alpha_k = \int^t d\tau \frac{\langle k|\partial H/\partial 
\tau |0\rangle}{\epsilon_k (\tau)} e^{ -i\int^t_{\tau} 
\epsilon_k (\tau') d \tau' } \, , 
\label{e9} \end{equation} 
where $\epsilon_k$ is the energy of the $k$th band relative to 
$\epsilon_0$. Since the total phases accumulated by the qubit 
states during the gate-voltage pulses are of the order of $\pi$, 
the adiabatic condition also means that the pulse amplitude 
should be small, i.e., $q_0=q^*+x(t)$, $x\ll 1$. 
Therefore, the energies and matrix elements in Eq.~(\ref{e9})
can be evaluated at $q_0=q^*$, and Eq.~(\ref{e9}) reduces to 
\begin{equation}
|\alpha_k| = 2 E_C |\langle k|(n-q)|0\rangle| |x(\epsilon_k)|\, . 
\label{e10} \end{equation} 
Here, $x(\epsilon_k)= \int d\tau x(\tau) e^{i \epsilon_k \tau }$ is
the Fourier component of $x(t)$ which decreases exponentially in
$\epsilon_k \tau_0$ for smooth pulses, where $\tau_0$ is the
characteristic pulse time. For $E_J\ll E_C$, the energy-band
gap of the transformer is of the order of $E_J$, and this sets
a lower limit for the pulse time through Eq.~(\ref{e10}). For 
$E_J \ge E_C$, the gap between the energy bands grows and the
adiabaticity condition is easier to satisfy (see the discussion after
Eq.~(\ref{e3}).

In conclusion, we have proposed a method to couple two Josephson
charge qubits by an electrostatic transformer that produces an
effective coupling that can be varied in magnitude and sign. We
have given an explicit expression of this coupling as a discretized
second derivative of the energy band of a small Josephson junction.
Our proposal 
works even for
asymmetric structures and in the presence of parasitic capacitances.
It allows the implementation of a variety of two-qubit
gates, and we have explicitly demonstrated how to build a phase gate.

C.~B. would like to thank the University of Stony Brook for its
hospitality during a one-month stay. 
The authors acknowledge 
discussions with K.K. Likharev which stimulated this work, and 
useful discussions of the results with J.E. Lukens and Yu.A. Pashkin. 
This work was supported in part by the NSF under grant \# 0121428, 
by ARDA and DOD under the DURINT grant \# F49620-01-1-0439 (D.V.A.), 
and by the Swiss NSF and the NCCR Nanoscience (C.~B.).


\begin{references}
\bibitem{b19} The geometric phase that in general could also be 
created by adiabatic evolution is absent in this case since 
only one parameter, $q_0$, affects the transformer 
wavefunction.
\end{references}
\end{document}